\documentclass[twocolumn,10pt]{asme2e}
\usepackage{graphicx}

\confshortname{ICONE25}
\conffullname{Proceedings of the 25th International Conference on Nuclear Engineering}

\confdate{May 14-18}
\confyear{2017}
\confcity{Shanghai}
\confcountry{China}

\papernum{ICONE25-66250 }

\title{Study of Minor Actinides Transmutation in PWR MOX fuel}

\author{Shengli Chen
    \affiliation{
	Sino-French Institute of \\
Nuclear Engineering and Technology\\
Sun Yat-Sen University\\
Zhuhai, 519082, Guangdong, China
    }	
}

\author{Cenxi Yuan\thanks{Address all correspondence to this author.}
    \affiliation{
    	Sino-French Institute of \\
Nuclear Engineering and Technology\\
Sun Yat-Sen University\\
Zhuhai, 519082, Guangdong, China\\
	yuancx@mail.sysu.edu.cn
    }
}
\author{Jingxia Wu
    \affiliation{
    	Sino-French Institute of \\
Nuclear Engineering and Technology\\
Sun Yat-Sen University\\
Zhuhai, 519082, Guangdong, China
    }
}

\author{Yaolei Zou
    \affiliation{
    	Sino-French Institute of \\
Nuclear Engineering and Technology\\
Sun Yat-Sen University\\
Zhuhai, 519082, Guangdong, China
    }
}

\begin{document}

\maketitle

\begin{abstract}
{\it The management of long-lived radionuclides in spent fuel is a key issue to achieve the closed nuclear fuel cycle and the sustainable development of nuclear energy. Partitioning-Transmutation is supposed to be an efficient method to treat the long-lived radionuclides in spent fuel. Some Minor Actinides (MAs) have very long half-lives among the radionuclides in the spent fuel. Accordingly, the study of MAs transmutation is a significant work for the post-processing of spent fuel.

In the present work, the transmutations in Pressurized Water Reactor (PWR) mixed oxide (MOX) fuel are investigated through the Monte Carlo based code RMC. Two kinds of MAs, $^{237}$Np and five MAs ($^{237}$Np, $^{241}$Am, $^{243}$Am, $^{244}$Cm and $^{245}$Cm) are incorporated homogeneously into the MOX fuel assembly. The transmutation of MAs is simulated with different initial MOX concentrations.

The results indicate an overall nice efficiency of transmutation in both initial MOX concentrations, especially for the two kinds of MAs primarily generated in the UOX fuel, $^{237}$Np and $^{241}$Am. In addition, the inclusion of $^{237}$Np in MOX has no large influence for other MAs, while the transmutation efficiency of $^{237}$Np is excellent.
The transmutation of MAs in MOX fuel depletion is expected to be a new, efficient nuclear spent fuel management method for the future nuclear power generation.}
\end{abstract}

\section{INTRODUCTION}

The management of long-lived radioactive production in the spent nuclear fuel is one of the most difficult issues associated with nuclear power generation. One of the concepts for dealing with the waste, so-called open cycle, is to solidify it and to confine it in deep geological formations over some hundreds of thousands years. However, it is very difficult to verify the assurance of confinement for such long time and the quantity of spent fuel is too large to be confined. Therefore, the closed cycle of spent fuel and the transmutation of some long-lived radioactive nuclides have been proposed as an alternative concept for solving the problem.

The uranium-plutonium Mixed OXide (MOX) fuel was studied and tested in the past decades in some western European countries and Japan \cite{Frank2001} in order to recycle the plutonium, which is about 1\% in the spent fuel \cite{IAEA2009} and has long half-life and large fission cross section for some fissionable nuclides. At present, MOX fuel is used in Pressurized Water Reactor (PWR), the most common commercial reactor around the world. MOX fuel is expected to be used more and more in order to reduce the spent fuel storage burden. This is the nuclear power generation trend of the Chinese utility companies; for example the MOX fuel will be used in Taishan Evolutionary Power Reactor (EPR) in the future.

The Minor Actinides (MAs) have very long decay half-lives among all radioactive nuclides in the spent fuel. Accordingly, the study of MAs transmutation is a significant work for the post-processing of spent fuel. The transmutation of MAs is studied in both PWR \cite{Liu2014,Hu2015}, Fast Breeder Reactor (FBR) \cite{Nishihara2010,Meiliza2008,Toshio2002,Hu2010} and subcritical reactors \cite{Beller2001,Herrera2007}. The transmutation of MAs in different reactors has been introduced in Ref.\cite{LIANG2017}. The majority of commercial nuclear power reactors operating in the world are PWRs. It is thus important to investigate the transmutation of MAs in the operating PWRs, which provides a potential approach to reduce the inventory of high level long-lived radioactive MAs in the world.

$^{237}$Np constitutes 56.2\% of the total MAs in the depleted nuclear fuel of PWRs \cite{Liu2014} and it has the longest decay half-life among MAs. Accordingly, it is also important to study the transmutation of $^{237}$Np loading in addition to the transmutation of a composition of MAs.

The transmutation of MAs in PWR has been previously investigated in Refs. \cite{Liu2014} and \cite{Hu2015}, which are based on UO$_{2}$ fuel assembly. Liu \emph{et~al.} \cite{Liu2014} focused in both homogeneous and heterogeneous MAs distribution in a reactor core. Hu \emph{et~al.} \cite{Hu2015} studied the heterogeneous core with MAs coated poison rods. The heterogeneous distribution is excellent in the consideration of spatial self-shielding and the reactivity control. In the present work, the transmutation of MAs is investigated in a MOX fuel based PWR, which assumes MAs homogeneous distributed in the fuel rods. It is necessary for the present study because of several reasons. Firstly, the neutronic properties of MAs are different in MOX fuel compared with those in UO$_{2}$ fuel. Secondly, the plutonium in MOX fuel and the MAs are extracted from the depleted nuclear fuel, which may simplify the post-processing of the depleted fuel that is not necessary to completely separate them. Lastly, the homogeneous distribution is easier for the fuel assembly fabrication.
	
In this study, the MAs are incorporated homogeneously into the MOX fuel rod. The present work studies the concentrations of some important MAs variated with the depletion of fuel. Low and high concentration MOX fuel in PWR are analyzed first. Then, we focus on 1\%wt MAs composition loading and 1\%wt $^{237}$Np loading, respectively. Finally, the effective multiplication factor k is compared among different cases.

\section{METHODOLOGY AND INPUT PARAMETERS}

\begin{figure}
\centering
\includegraphics[width=1.0\linewidth]{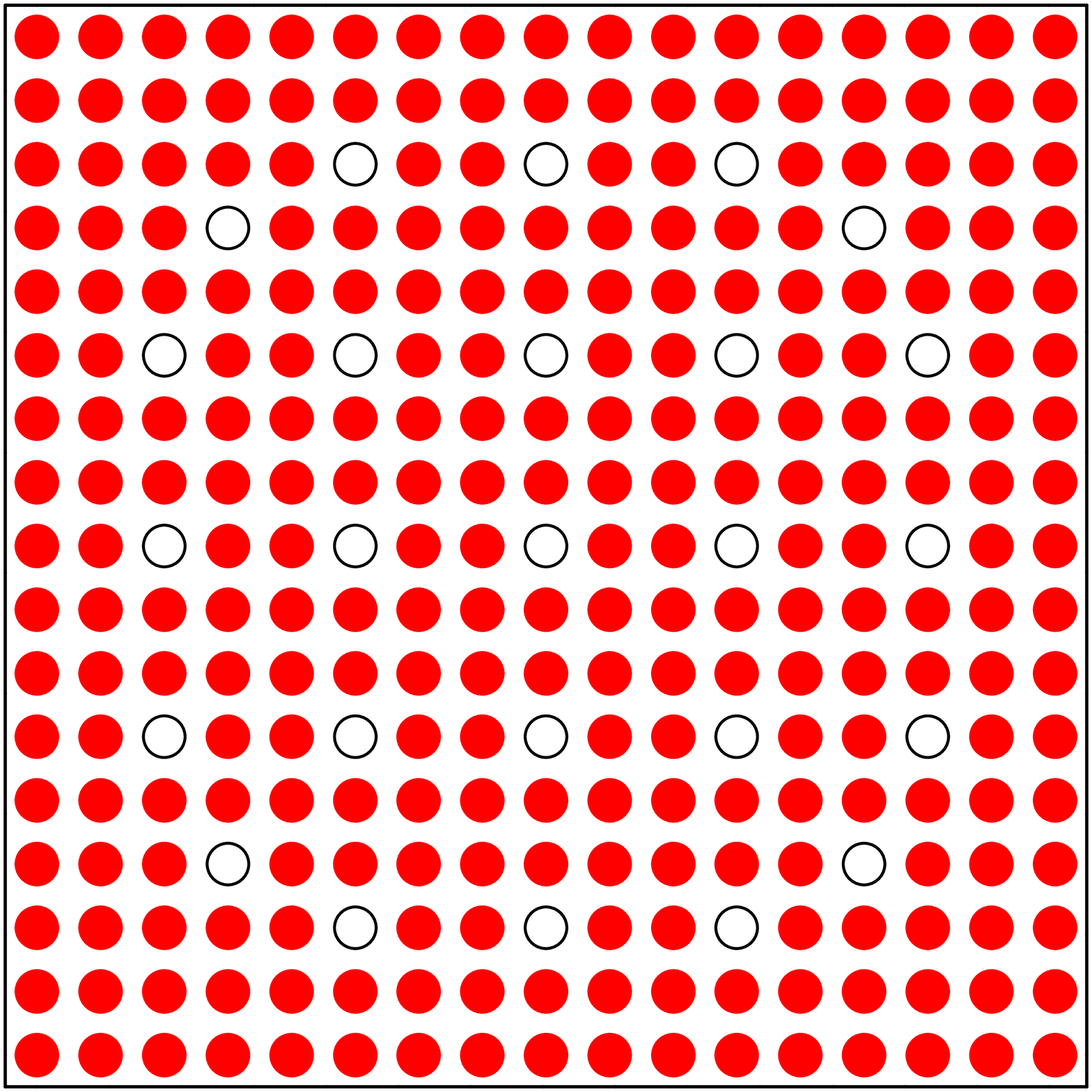}
\caption{17{$\times$}17 PWR lattice configuration}
\label{fig1}
\end{figure}

The 17{$\times$}17 PWR assembly design is used in the present study, as shown in Fig.\ref{fig1}. The larger rings placed within the lattice represent locations of guide tubes and instrumentation tube. When there is no insertion of control rods or instrument, these tubes are full of moderator, which is the case for present work. All simulations are carried out on an assembly. The parameters are shown in Table \ref{table1}, the same as listed in Ref.\cite{Chen2017}, are shown in Table 1. These simulations are performed using the RMC Monte Carlo code \cite{Wang2013}, which is a 3-D Monte Carlo neutron transport code developed by Tsinghua University. The RMC is able to deal with complex geometry and use continuous energy pointwise ENDF/B-VII.0-based cross sections for different materials and at different temperatures \cite{Li2011}. It has both criticality and burnup calculations, which can obtain the effective multiplication factor and the nuclide concentrations at different burnup level. The data libraries of the present RMC code are processed by the code RXSP \cite{Yu2013}
	
Three random number generotors are contained in the RMC code. The code supports also the parallel critical calculation, adopts the asymptotical super-history method and asymptotical Wielandt method to accelerate the source convergence, reducing the inactive generations, which can reduce the calculation time, while the time-consuming problem is one of the most disadvantages of Monte Carlo method. In the RMC, the fission source iteration method is adopted in the critical calculation. It contains the Shannon Entropy statistics, and the fission matrix statistics.

About 1500 nuclides are contained in the burnup calculation, which is realized in the embedded DEPTH code \cite{She2013,She2013w}. The depletion calculation is the resolution of the Bateman equation. The RMC firstly calculates the data such as neutron flux and one-group reaction cross-sections, by the module of critical calculation (continuous-energy), and then passing these data to DEPTH. Secondly, DEPTH completes the point burnup calculation to acquire the new nuclides density, and then passing them to the module of critical calculation. Moreover, it integrates also the latest point burnup libraries of ORIGEN-S and ORIGEN-2 \cite{Parks}. In the present work, the depletion is calculated with steps of 0.33 (1.33 and 1.67 resp.) MWd/kg for the first (second and third resp.) 10 steps, and 3.33MWd/kg in the following steps.


\begin{table*}
	\centering
	\caption{\label{table1}Geometry, material, and model specification }
	\begin{tabular}{ccc}
		\hline
		Property & Unit & Value  \\
		\hline
		Assembly fuel height & cm & 365.76 \\
		\hline
		Clad composition & wt\% & Zr-4: Fe/Cr/Zr/Sn = 0.15/0.1/98.26/1.49 \\
		\hline
		Fuel pellet radius & mm & 4.096 \\
		\hline
		Gap thickness & {$\mu$}m & 82.55 \\
		\hline
		Cladding Inner Radius & mm &  4.178 \\
		\hline
		Cladding thickness & {$\mu$}m & 571.5 \\
		\hline
		Cladding Outer Radius & mm & 4.750 \\
		\hline
		Pitch-to-Diameter ratio  &   & 1.326 \\
		\hline
		Cladding IR of guide tube &	mm	& 5.690 \\
		\hline
		Cladding OR of guide tube &	mm	& 6.147 \\
		\hline
		Fuel density \cite{Popov2000} & g/{$cm^3$} & Low MOX: 10.993  \\
        \hline
		Fuel density \cite{Popov2000} & g/{$cm^3$} & High MOX: 11.026 \\
		\hline
		Coolant density	& g/{$cm^3$}	& 0.7119 \\
		\hline
		Helium density	& g/L	& 1.625 (2.0 MPa) \\
		\hline
		Cladding density	& g/{$cm^3$}	& 6.56 \\
		\hline
		Coolant temperature	& K	& 574 \\
		\hline
		Fuel temperature	& K	& 924 \\
		\hline
		Clad and gap temperature & K	& 574 \\
		\hline
		Boron concentration	& ppm	& 630 \\
		\hline
		Moderator-to-fuel ratio	& &	2 \\
		\hline
		Boundary conditions	& &	Reflective  \\
		\hline
	\end{tabular}
\end{table*}

Monte Carlo method is widely used in various areas besides reactor physics. Taking nuclear science for example, fundamental properties of nuclei can be solved with traditional nuclear shell model \cite{Yuan2012,Yuan2014,Yuan2016} or Monte Carlo shell model \cite{Otsuka2001}. One of the main advantages of Monte Carlo method is that the simulation can be achieved in any complex geometry. In addition, there are less assumptions and approximations with Monte Carlo method than deterministic method. As a consequent, Monte Carlo based codes can give more precise results. However, in order to obtain the results with small variance, much more simulation time is needed with Monte Carlo method. An advanced method is used in RMC and it reduces much simulation time compared with other Monte Carlo based codes.

Two sets of MOX fuel are considered in the present work in order to compare the transmutation efficiency with different MOX concentration. The high (or low) concentration MOX fuel contains 9.8\% (or 4.2\%) Pu isotopes in the fuel and the enrichment of each isotope is shown in Table \ref{table2}.

\begin{table*}
	\centering
		\caption{\label{table2}Actinide initial enrichment (\%) of high and low concentration MOX fuel}
	\begin{tabular}{cccccccc}
		\hline
		 & $^{235}$U &	$^{238}$U	& $^{238}$Pu	& $^{239}$Pu	& $^{240}$Pu	& $^{241}$Pu	& $^{242}$Pu \\
		\hline
		Low MOX	& 0.24	& 95.562	& 0.063	& 2.528	& 1.029	& 0.37	& 0.21  \\
		\hline
		High MOX &	0.226 &	89.975	& 0.147	& 5.9	& 2.401	& 0.862	& 0.49 \\
		\hline
	\end{tabular}
\end{table*}

In general, the decay half-life of MA is very long. Among MAs nuclides, the most enrich nuclide in the spent fuel $^{237}$Np has the longest decay half-life 2.14{$\times 10^6$} years. After $^{237}$Np, $^{245}$Cm and $^{243}$Am have the longest decay half-life, which is 8500 years and 7070 years respectively. Another enriched nuclide in the spent fuel is $^{241}$Am, which half-life is 432 years. However, this is not the case for $^{242}$Am due to its high thermal neutron fission cross section and its short half-life (16 hours). It is thus not necessary to transmute $^{242}$Am in the MOX fuel.

Due to the long half-live of $^{237}$Np, its transmutation is of large importance for the post-processing of spent fuel. In this context, loading of 1\% $^{237}$Np is studied in the present work. In addition, in order to transmute other long-live MAs, it is also of great interest to study the mixed long-live MAs. In the case of 1\%wt mixed MAs loading in two concentration MOX fuel, the percentage of each MA nuclide is shown in Table \ref{table3}. This corresponds to the results of a 3GW thermal power reactor after 10 years cooling \cite{Broeders2000}.

\begin{table}
	\centering
		\caption{\label{table3}Percentage of loading MAs in MOX}
	\begin{tabular}{cccccc}
		\hline
		 Isotope &	$^{237}$Np &	$^{241}$Am	& $^{243}$Am	& $^{244}$Cm	& $^{245}$Cm \\
		\hline
		wt\% &	41.80 &	47.86	& 8.62	& 1.63	& 0.09  \\
		\hline
	\end{tabular}
\end{table}

\section{RESULTS AND DISCUSSIONS}

\begin{figure}
\centering
\includegraphics[width=1.0\linewidth]{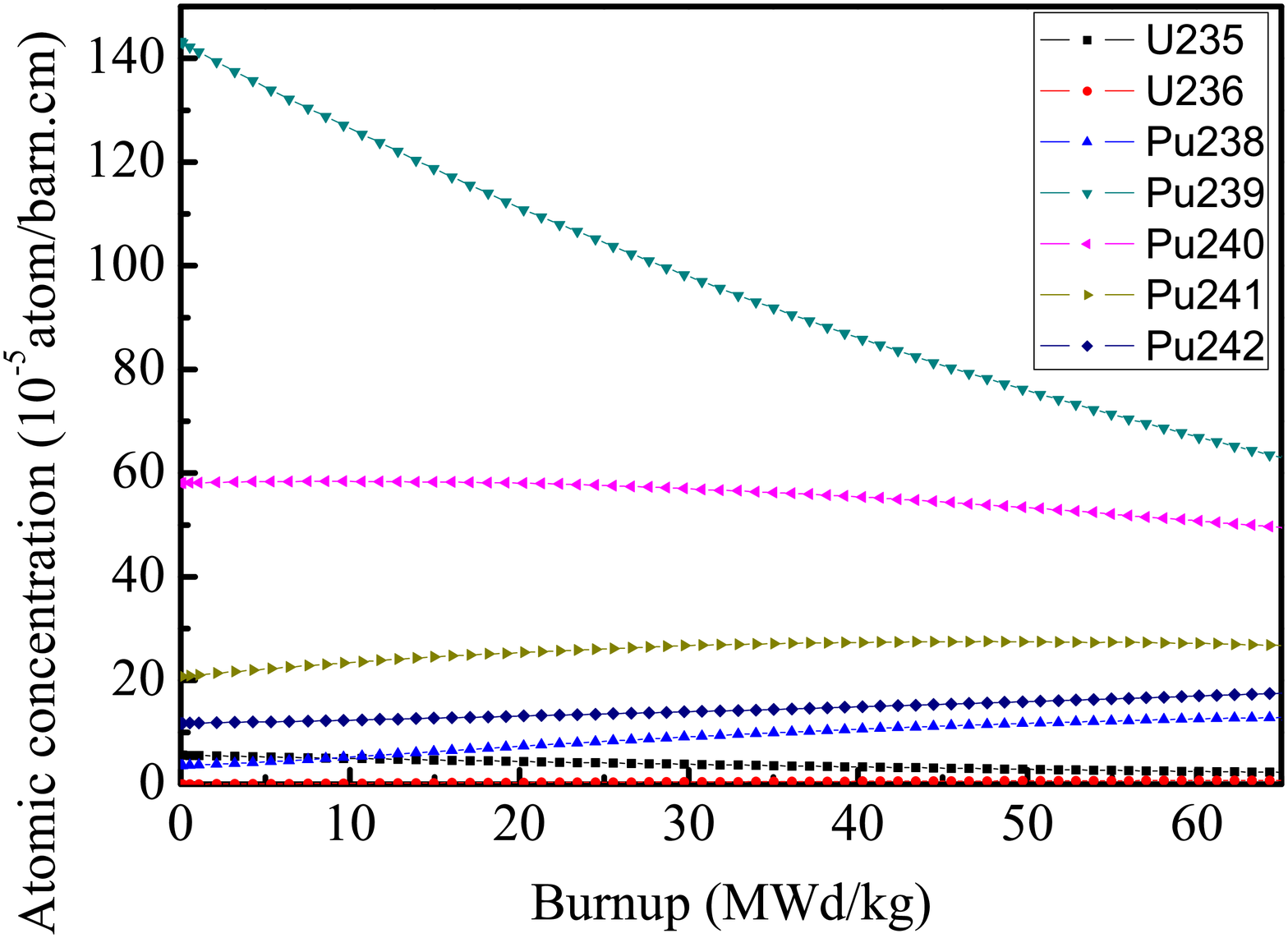}
\includegraphics[width=1.0\linewidth]{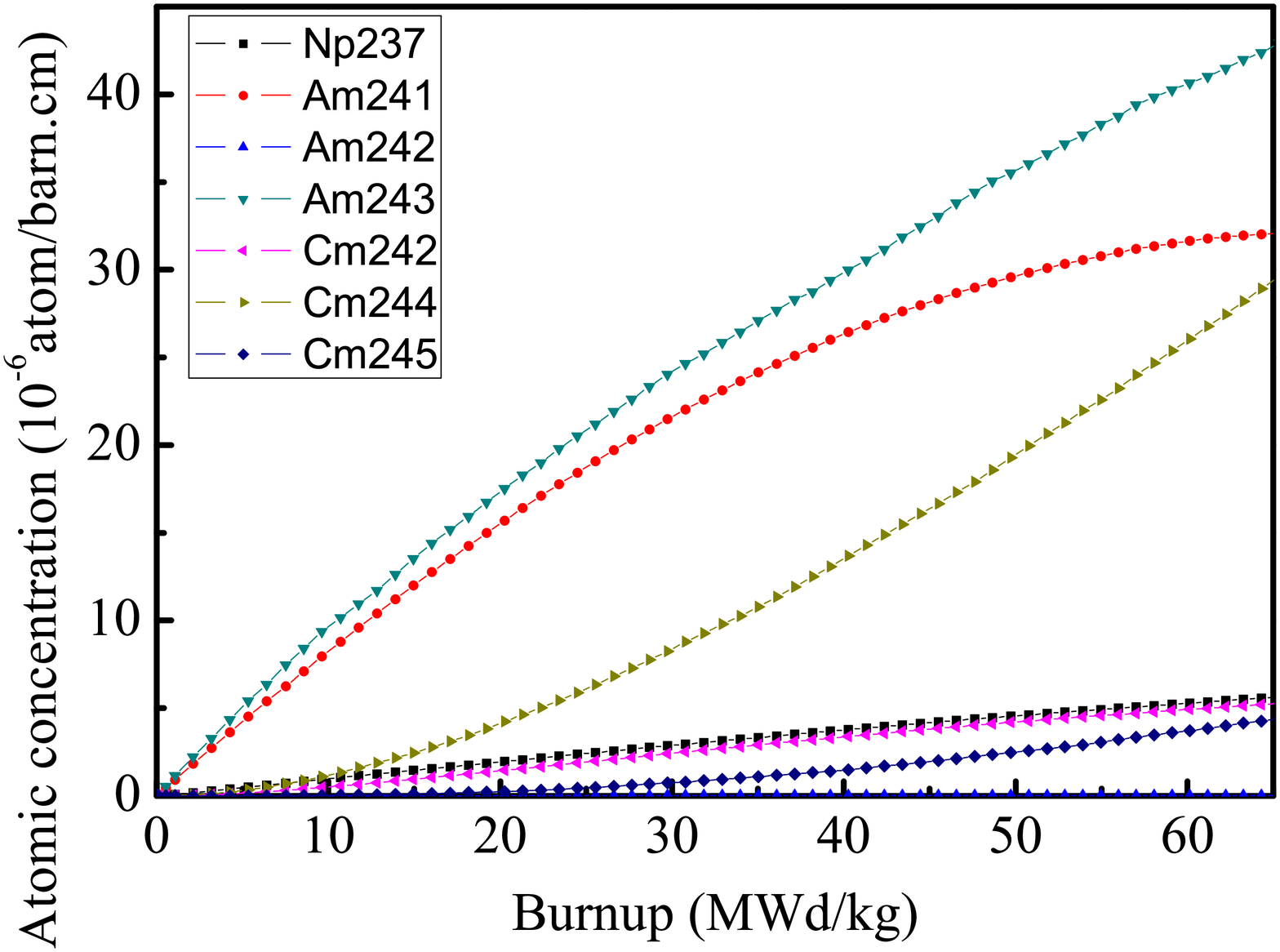}
\caption{Actinide concentration of high concentration MOX}
\label{fig2}
\end{figure}

\begin{figure}
\centering
\includegraphics[width=1.0\linewidth]{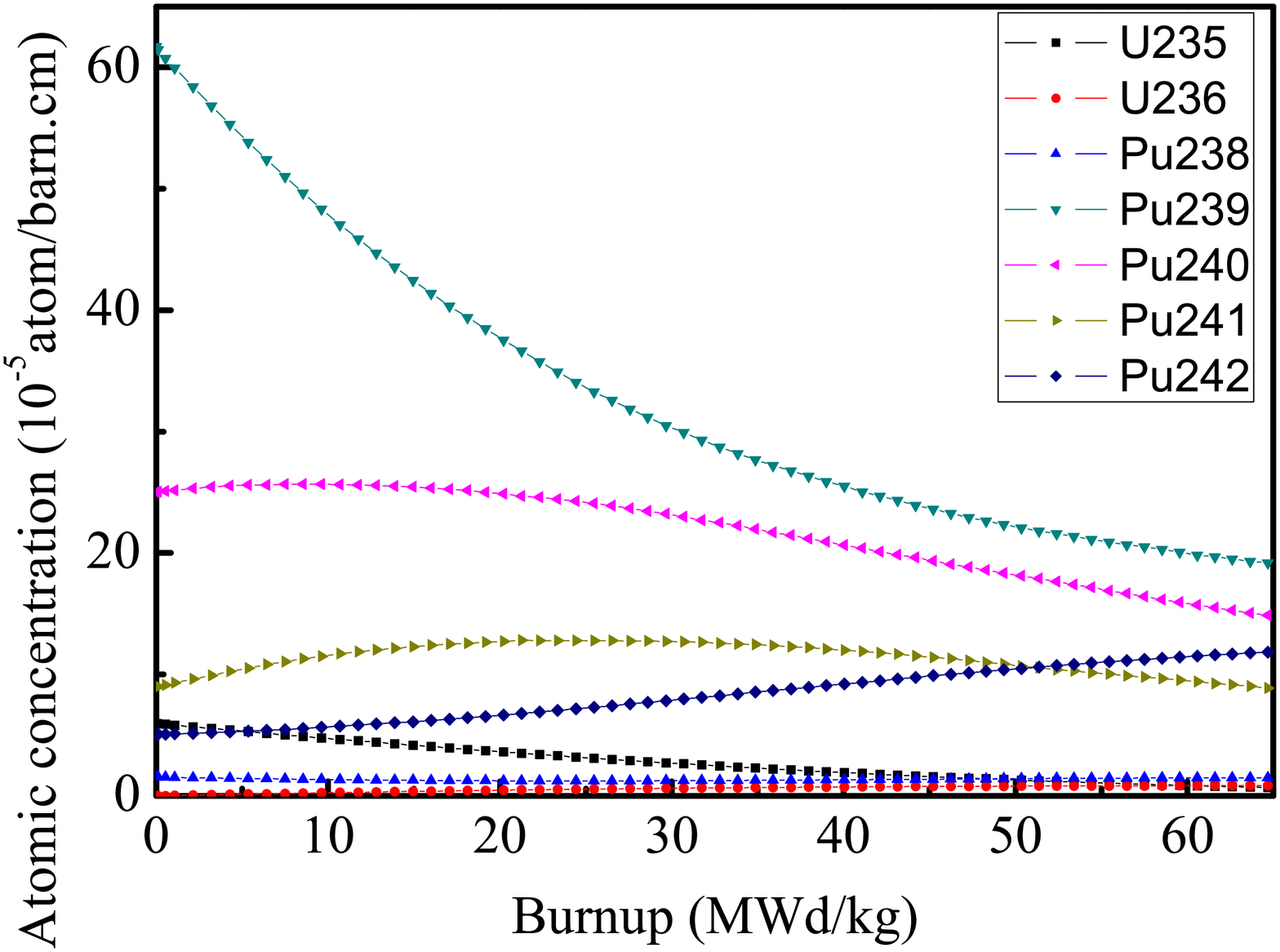}
\includegraphics[width=1.0\linewidth]{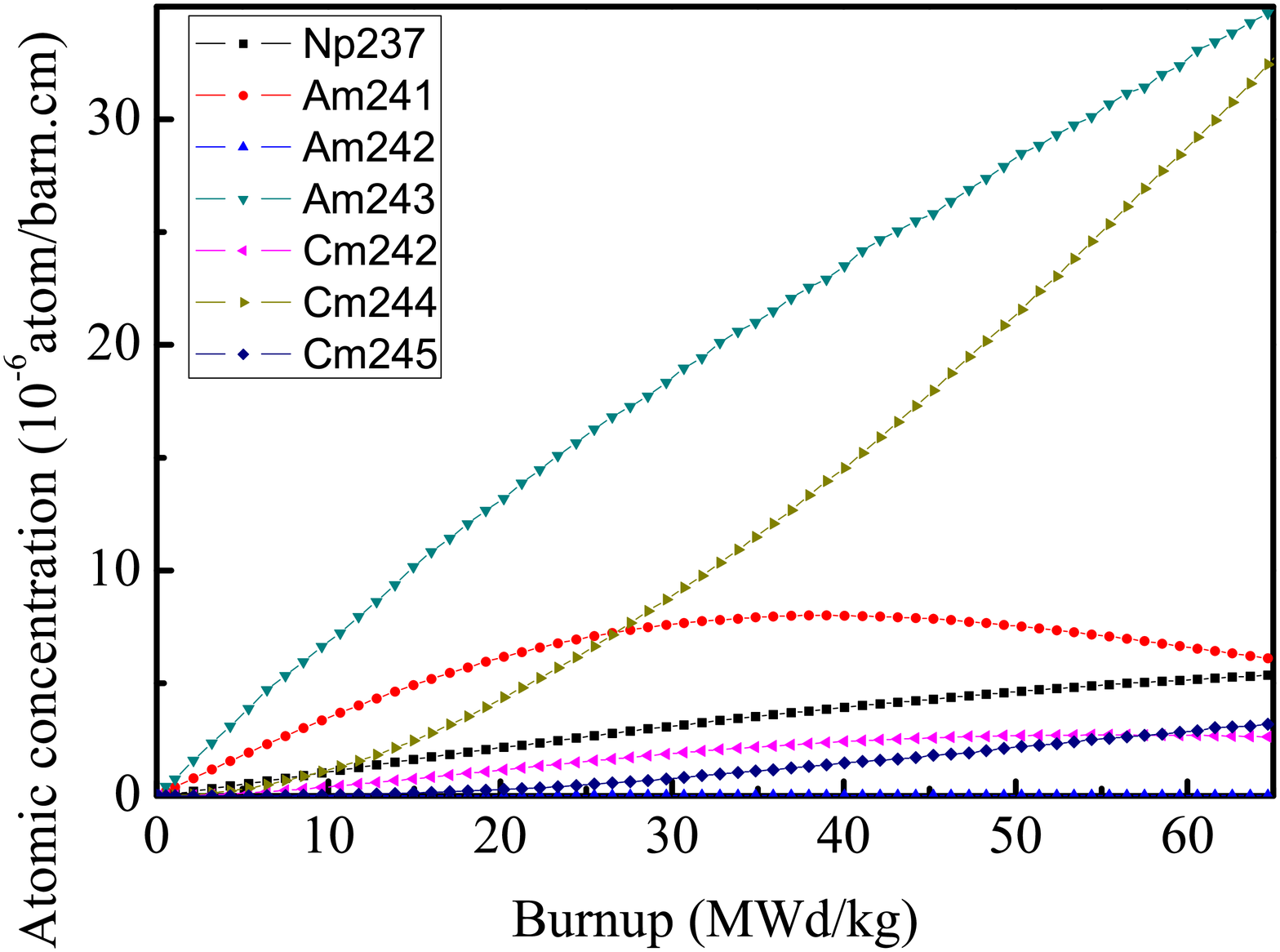}
\caption{Actinide concentration of low concentration MOX}
\label{fig3}
\end{figure}

The nuclide concentrations of two different sets of MOX fuel as the function of burnup are shown in Fig.\ref{fig2} and Fig.\ref{fig3}. The results are similar to the results obtained in Ref. \cite{Yuan2016c}, which is calculated in a single fuel rod.

In general, the produced MAs in the low concentration MOX fuel are less than that in the high concentration MOX fuel because of the less initial concentration of plutonium, which produces MAs nuclides through a series of neutron capture reactions and {$\beta$} decays. The most evidently difference is that the $^{241}$Am produced in low concentration MOX is much less than that in high concentration MOX. The reason is that $^{241}$Am nuclei are mostly from the {$\beta$} decay of $^{241}$Pu, which has much lower concentration in low concentration MOX. In this case, both the concentrations of $^{241}$Pu and $^{241}$Am decrease at high burnup level. For $^{241}$Am, although its cross section of thermal neutron fission is only 2.2 barn, its cross section of thermal neutron capture is 412.3 barn \cite{Iwasaki2002}. At high burnup level, the increment of $^{241}$Am coming from $^{241}$Pu is not enough to compensate the decrement of $^{241}$Am through (n,{$\gamma$}) reaction. $^{241}$Am can also decay through emitting an {$\alpha$} particle, its long half-life (432.6 year) ensures that {$\alpha$} decay rarely affects its concentration during the fuel rod loading in PWR. The decrement of $^{242}$Cm in low concentration MOX is similar to that of $^{241}$Am. There is no large difference of $^{235}$U and corresponding $^{237}$Np, which is produced by two times of (n,{$\gamma$}) reactions and one time of {$\beta$} decay.

In both high and low concentration MOX fuel, concentrations of $^{243}$Am and $^{244}$Cm increase rapidly with burnup. This is due to their production rates are much higher than their transmutation rates. Both of them have very low fission cross sections \cite{Iwasaki2002} because of their non-fissile nature from even numbers of neutrons. Although their capture cross sections is not low \cite{Iwasaki2002}, their transmutation rates cannot overcome the corresponding production rates. It is because that $^{243}$Am is the production of $^{242}$Pu by one time (n,{$\gamma$}) reaction and one time of {$\beta$} decay. The concentration of $^{242}$Pu is much larger than that of $^{243}$Am. The capture cross section of $^{242}$Pu is also large. In the case of $^{244}$Cm, it is produced by one time (n,{$\gamma$}) reaction and one time of {$\beta$} decay of $^{243}$Am, of which the initial concentration is zero. The increment rate of $^{243}$Am is much larger than that of $^{244}$Cm at the beginning of the cycle. On the contrary, the faster increment rate of $^{244}$Cm than that of $^{243}$Am at high burnup level is because that the capture cross section of $^{244}$Cm is much lower than that of $^{243}$Am.
	
\begin{figure}
\centering
\includegraphics[width=1.0\linewidth]{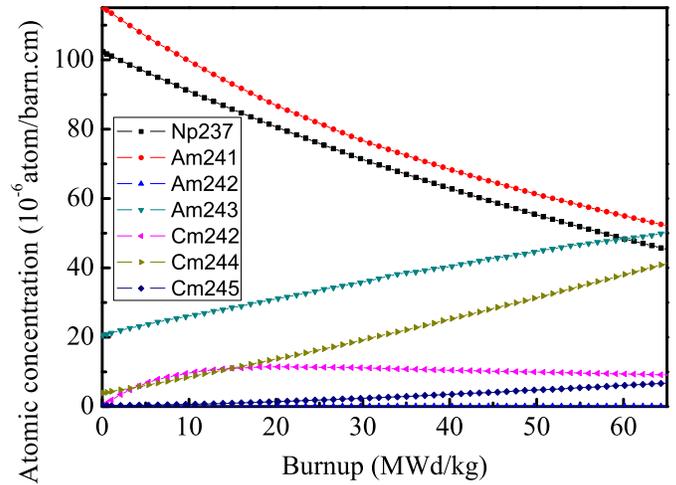}
\caption{MA concentration of high concentration MOX with 1\% MAs loading}
\label{fig4}
\end{figure}

\begin{figure}
\centering
\includegraphics[width=1.0\linewidth]{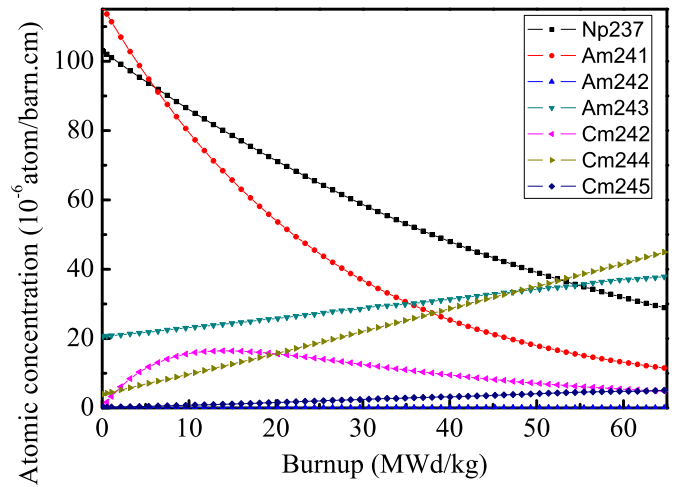}
\caption{MA concentration of low concentration MOX with 1\% MAs loading}
\label{fig5}
\end{figure}

If the MAs are loaded in MOX fuel, as discussed in the last section, the transmutation efficiency of MAs in low concentration MOX fuel is generally better due to the lower initial plutonium concentration, especially for the $^{241}$Am (as shown in Fig.\ref{fig4} and Fig.\ref{fig5}). This is not surprising because of the competition between Pu and MAs on absorbing neutrons. Generally speaking, high concentration of Pu suppresses the number of neutrons absorbed by MAs and corresponding reaction rates for transmutation which directly influence the concentrations of MAs. However, such mechanism has opposite effect on certain nuclide, such as $^{244}$Cm, which is slightly sparser in high concentration MOX fuel. Because $^{244}$Cm mostly comes from $^{243}$Am through one time of neutron capture and one time of {$\beta$} decay. The (n,{$\gamma$}) reaction of $^{243}$Am and corresponding production rate of $^{244}$Cm are suppressed in high concentration MOX fuel.
	
$^{241}$Am has almost the same transmutation rate as $^{237}$Np in high concentration MOX, but it has larger transmutation rate than the latter in low concentration MOX fuel. This is because less $^{241}$Am is produced due to less concentration of Pu isotopes, especially $^{241}$Pu, in low concentration MOX.

\begin{figure}
\centering
\includegraphics[width=1.0\linewidth]{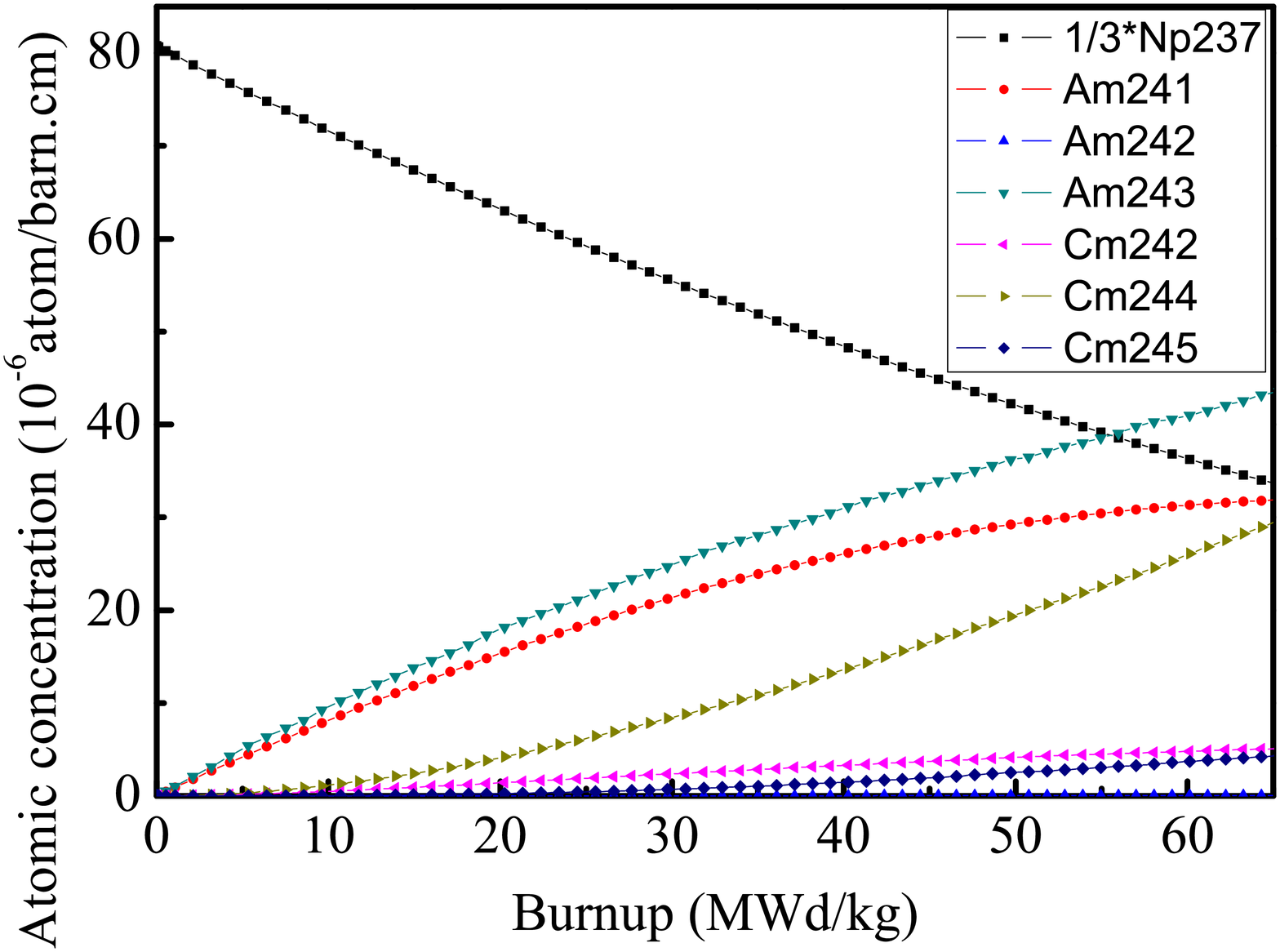}
\caption{MA concentration of high concentration MOX with 1\% $^{237}$Np loading}
\label{fig6}
\end{figure}

\begin{figure}
\centering
\includegraphics[width=1.0\linewidth]{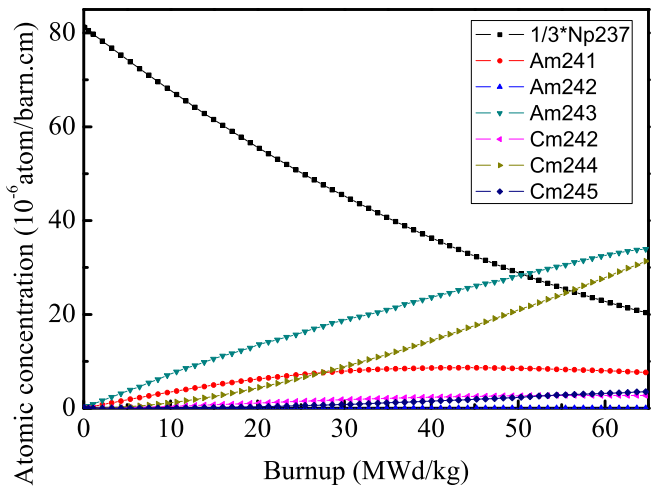}
\caption{MA concentration of low concentration MOX with 1\% $^{237}$Np loading}
\label{fig7}
\end{figure}

Compared with cases without loading MAs, it is remarkable that there are no large variations on nuclide concentrations of other MAs except $^{237}$Np and $^{241}$Am after 1\% MAs is loaded. The transmutation rates of these two nuclides are around 55\% for both in high concentration MOX and 72\% for $^{237}$Np and 90\% for $^{241}$Am in low concentration MOX when the burnup is up to 64 GWd/t. In the high concentration MOX fuel, it should be also noted that at 64 GWd/t burnup level, about 6.3 {$\times 10^{19} at./cm^3$}  of $^{241}$Am has been transmuted and only around 2.0{$\times 10^{19} at./cm^3$} of $^{241}$Am is produced compared with the case without loading MAs.
	
In the case that only 1\% $^{237}$Np is loaded (Fig.\ref{fig6} and Fig.\ref{fig7}), there are no large influence on other MAs. The transmutation rate in high concentration MOX is lower than that in low concentration MOX. There are a little fewer $^{235}$U in high concentration MOX, which produces the $^{237}$Np by two times of (n,{$\gamma$}) reactions and one time of {$\beta$} decay. The reason is discussed before that high concentration of Pu suppresses the number of neutrons absorbed by $^{237}$Np and corresponding absorption rate for $^{237}$Np Similar to the previous discussions, the transmutation rate of $^{237}$Np is larger in low concentration MOX fuel, which is 75\%, while that in high concentration MOX is 58\%.
The present work shows the excellent transmutation efficiency of $^{237}$Np, which half-life is 2.14{$\times 10^6$} years. Taking the lowest transmutation rate 55\% in high concentration MOX at 64 GWd/t burnup level for example, it is equivalent to 2.47{$\times 10^6$} years of natural decay. In addition, the production of $^{237}$Np still exists even though the MAs are not added in MOX or other nuclear fuels.

Transmutation is an excellent method to reduce the long-lived MAs and fission products. However, because the most long-lived MAs have large thermal neutron capture cross section but small thermal fission cross section, such as $^{237}$Np, $^{241}$Am and $^{243}$Am. Hence, it is important to study the effective multiplication k after loading the MAs.

\begin{figure}
\centering
\includegraphics[width=1.0\linewidth]{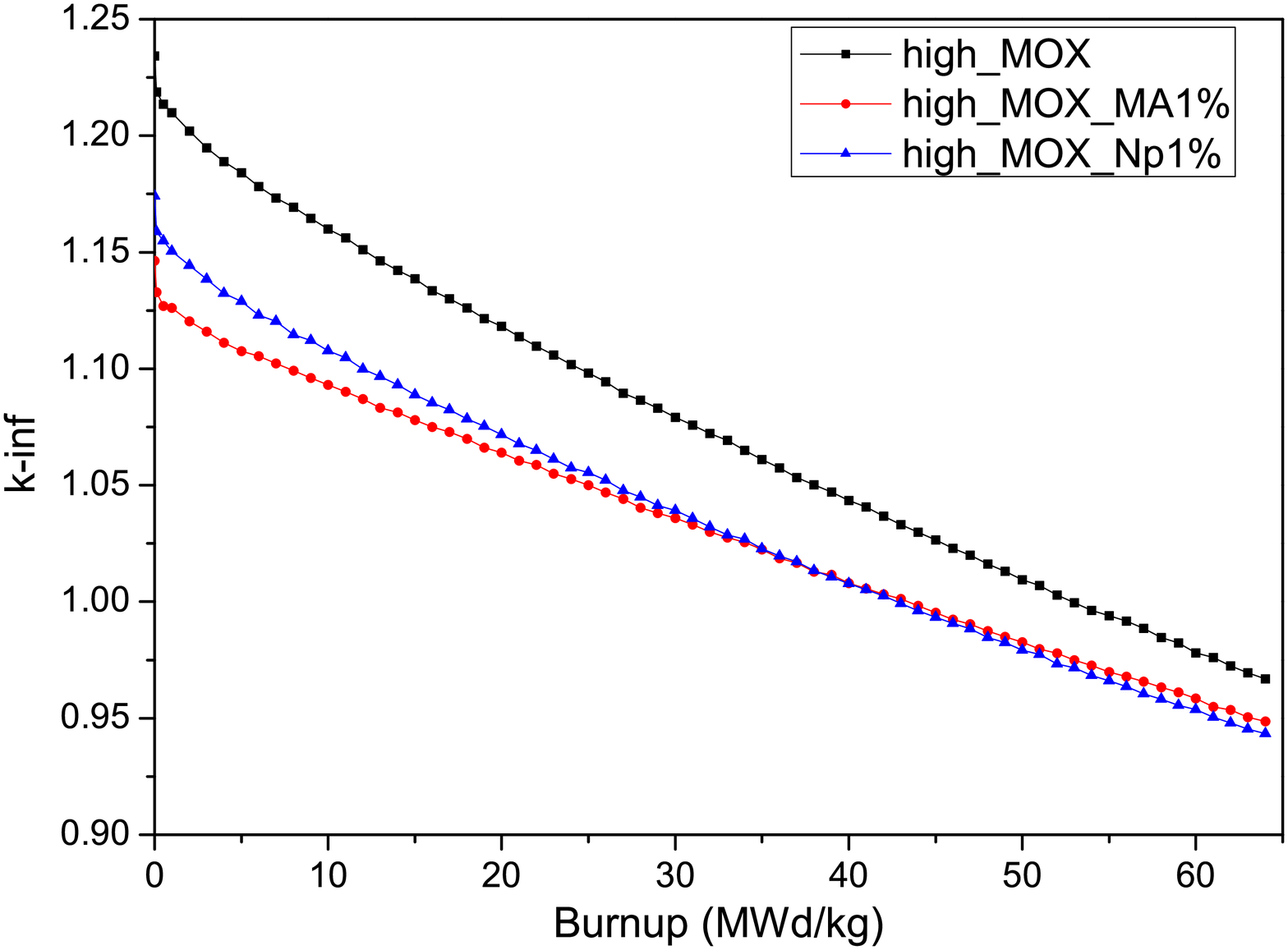}
\includegraphics[width=1.0\linewidth]{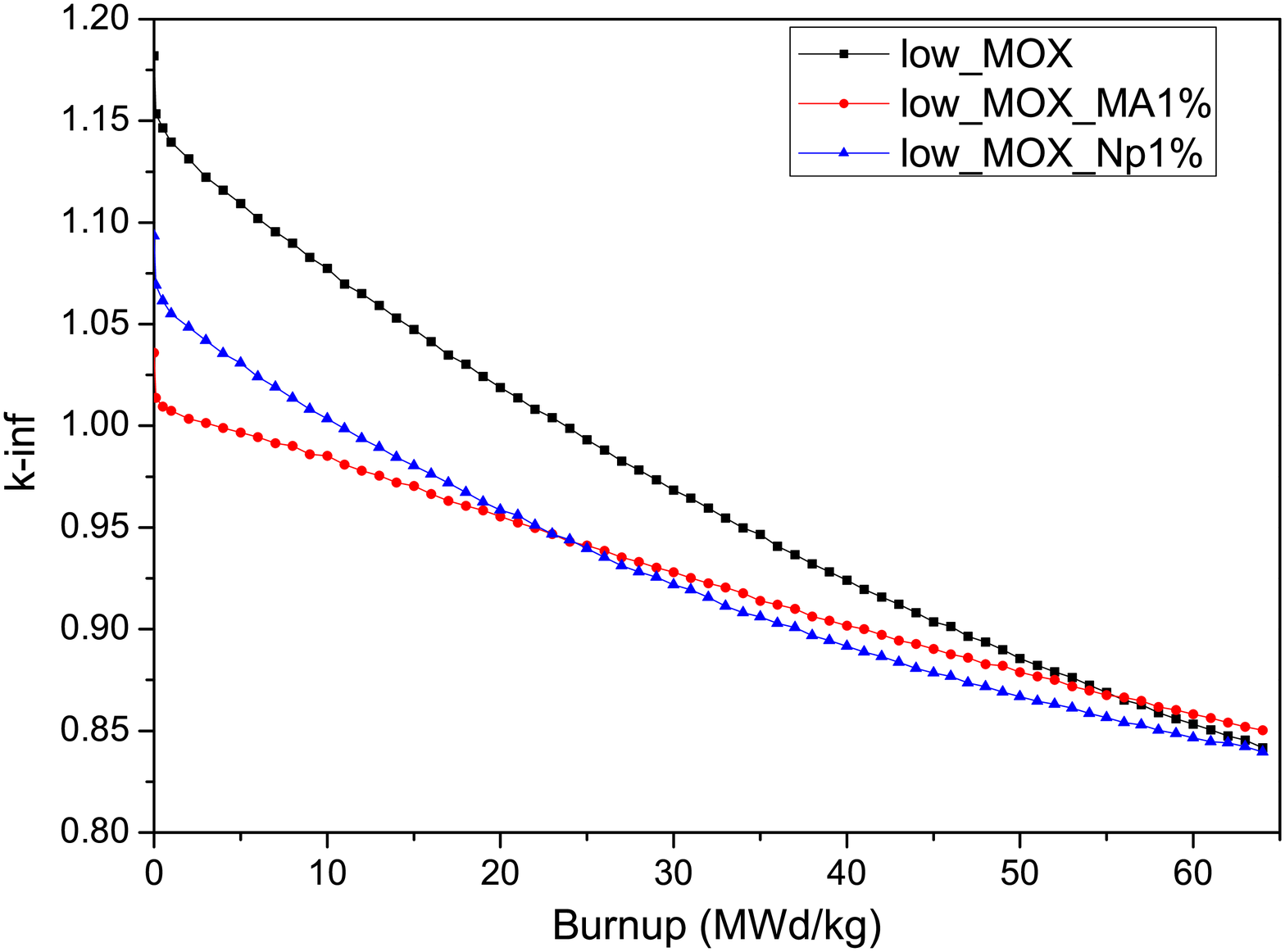}
\caption{The effective multiplication factor k for different cases}
\label{fig8}
\end{figure}

As shown in Fig.\ref{fig8}, the addition of MAs induces the large decrement of the reactivity because of the high thermal neutron capture cross section of the majority of added MAs. Initially, the decrement of the reactivity is more significant due to much higher thermal neutron cross section of $^{241}$Am than that of $^{237}$Np \cite{Iwasaki2002}.

However, in the case of 1\% MAs loading the factor k becomes larger than that in the 1\% $^{237}$Np loading case. This is due to the neutron absorption of $^{241}$Am, of which the production are $^{242}$Am and $^{242}${$^m$}Am. $^{242}$Am and $^{242}${$^m$}Am are both fissile nuclides and their thermal fission cross section are 1332 barn and 4899 barn, respectively \cite{Iwasaki2002}.

Many studies prove the rim effect that the reaction ratio is much higher around the periphery of the fuel rod \cite{Yuan2016w,Yuan2016ch}. The reduction of reactivity by MAs loading is also proved in this work and in Ref \cite{Liu2014}. Accordingly, it is also reasonable to transmute the MAs by heterogeneous addition, only in outer layer of the nuclear fuel. This can not only transmute the MAs nuclides, but also flatten the radial power distribution in a fuel rod.

In addition, as shown in Fig.\ref{fig8}, the difference of the reactivity between the MAs loading and normal MOX fuel decreases with the burnup due to the existence of some nuclides, which have large thermal neutron fission cross section, such as $^{242}$Am, $^{244}$Am and $^{245}$Cm. This indicates that MAs can be also used as burnable neutronic poison.

The present work has studied the transmutation of MAs with 1\% $^{237}$Np and 1\% mixed MAs loading in two sets of MOX fuel. The study in different moderator-to-fuel ratio cases is also interesting to analysis the transmutation efficiency in different nuclear core. Moreover, it is also important to study different percentage of MAs loading in order to maximize the transmutation efficiency in case of few reduction of the current cycle length. The sensitive and uncertainty analyses from nuclear data are also very important, such as an uncertainty analysis on the simple nuclear mass model, liquid drop model, is performed recently \cite{Yuan2016u}.

\section{CONCLUSION}

In summary, the present work has studied the transmutation of MAs in different MAs nuclides loading and in different concentration MOX using the Monte-Carlo based code RMC. It is proved that transmutation efficiency of MAs in PWR with MOX fuel is possible from a neutronic view. Our calculation results show the excellent transmutation efficiency of MAs nuclides, especially for $^{237}$Np when compared to natural decay, in both high and low concentration MOX fuel. In general, the low concentration MOX fuel has better transmutation efficiency due to lower initial Pu isotopic concentrations and smaller competence between Pu and MAs. However, attention of the large decrement of reactivity should also be paid, especially for the low concentration MOX fuel because it has less reactivity due to less fissile nuclides. The MA transmutation in MOX fuel depletion can be expected to be a new, efficient nuclear spent fuel management method for the future nuclear power generation. Further studies are needed for the realization of MAs transmutation in PWR.

\bibliographystyle{asmems4}

\begin{acknowledgment}
The authors acknowledge the authorized usage of the RMC code from Tsinghua University for this study. This work has been supported by the National Natural Science Foundation of China under Grant No. 11305272, the Special Program for Applied Research on Super Computation of the NSFC Guangdong Joint Fund (the second phase), the Guangdong Natural Science Foundation under Grant No. 2014A030313217, and the Pearl River S\&T Nova Program of Guangzhou under Grant No. 201506010060.
\end{acknowledgment}

%
% ...

\bibliography{asme2e}

\begin{thebibliography}{99}
\bibitem{Frank2001}
Frank N. von Hippel, 2001. Plutonium and Reprocessing of Spent Nuclear Fuel,Science, 293, 2397-2398.

\bibitem{IAEA2009}
IAEA, 2009. Status of minor actinide fuel development, ISSN 1995-7807, ISBN 978-92-0-106909-2

\bibitem{Liu2014}
Liu Bin, Wang Kai, Tu Jing, \emph{et~al.}, 2014. Transmutation of minor actinides in the pressurized water reactors, Annals of Nuclear Energy, 64, 86-92

\bibitem{Hu2015}
Hu Wenchao, Liu Bin, Ouyang Xiaoping, \emph{et~al.}, 2015. Minor actinide transmutation on PWR burnable poison rods, Annals of Nuclear Energy, 77, 74-82

\bibitem{Nishihara2010}
Nishihara K, Oigawa H, Nakayama S, \emph{et~al.}, 2010. Impact of partitioning and transmutation on high-level waste disposal for the fast breeder reactor fuel cycle, Journal of Nuclear Science and Technology, 47, 1101-1117

\bibitem{Meiliza2008}
Meiliza Y, Saito M, Sagara H. 2008. Protected plutonium breeding by transmutation of minor actinides in fast breeder reactor, Journal of Nuclear Science and Technology, 45, 230-237

\bibitem{Toshio2002}
Toshio Wakabayashi, 2002. Transmutation Characteristics of MA and LLFP in a Fast Reactor, Progress in Nuclear Energy, 40, 457-463

\bibitem{Hu2010}
HU Yun,WAN Kan,XU Mi. 2010. Transmutation of MA Nuclides in Sodium Cooled MOX Fuel Fast. Nuclear Power Engineering.

\bibitem{Beller2001}
Beller, D.E., Van Tuyle, G.J., Bennett, D., 2001. The U.S. accelerator transmutation of waste program. Nuclear Instruments and Methods in Physics Research A, 463, 468-486

\bibitem{Herrera2007}
Herrera-Martnez, A., Kadi, Y., Parks, G., 2007. Transmutation of nuclear waste in accelerator-driven systems: Thermal spectrum. Annals of Nuclear Energy 34, 550-563

\bibitem{LIANG2017}
LIANG Tongxiang and TANG Chunhe. 2003. Transmutation of long-lived nuclides. Nuclear Techniques, 26, 12

\bibitem{Chen2017}
Shengli Chen and Cenxi Yuan. Neutronic Analysis on Potential Accident Tolerant Fuel-Cladding Combination U{$_3$}Si{$_2$}-FeCrAl. Science and Technology of Nuclear Installations, 2017 (2017) 3146985

\bibitem{Wang2013}
K. Wang, Z. Li, D. She, \emph{et~al.}, RMC - A  Monte Carlo Code  for  Reactor  Physics  Analysis, International Conference on Mathematics and Computational Methods Applied to Nuclear Science and Engineering, M and C 2013, 1, p 89-104, Sun Valley, ID, USA, May 5-9, 2013.

\bibitem{Li2011}
Z. Li, K. Wang, X. Zhang. Research on Applying Neutron Transport Monte Carlo Method in Materials with Continuously Varying Cross-sections, M\&C 2011, Rio de Janeiro, RJ, Brazil, May 8-12, 2011.

\bibitem{Yu2013}
J. Yu, S. Li, K. Wang, \emph{et~al.}, The Development and Validation of Nuclear Cross Section Processing Code for Reactor-RXSP. The 2013 21st INTERNATIONAL CONFERENCE ON NUCLEAR ENGINEERING, July 29-August 2, 2013, Chengdu, China.

\bibitem{She2013}
D. She, Y. Liu, K. Wang, \emph{et~al.}, Development of Burnup Methods and Capabilities in Monte Carlo Code RMC. Annals of
Nuclear Energy, January 2013, 51: 289¨C294

\bibitem{She2013w}
D. She, K. Wang, G. Yu, Development of the point-depletion code DEPTH. Nuclear Engineering and Design, May, 2013,
258: 235¨C240

\bibitem{Parks}
C. V. Parks. Overview of ORIGEN2 and ORIGEN-S: Capabilities and Limitations

\bibitem{Yuan2012}
C. Yuan, T. Suzuki, T. Otsuka, \emph{et~al.}, Study of B, C, N, and O isotopes based on VMU, Physical Review C, 85, 064324 (2012).

\bibitem{Yuan2014}
C. Yuan, C. Qi, F. Xu, \emph{et~al.}, Mirror energy difference and the structure of loosely bound proton-rich nuclei around A=20, Physical Review C, 89, 044327 (2014).

\bibitem{Yuan2016}
C. Yuan, Z. Liu, F. Xu, \emph{et~al.}, Isomerism in the south-east of 132Sn and a predicted neutron-decaying isomer in 129Pd, Physics Letters B, 762 (2016) 237-242.

\bibitem{Otsuka2001}
T. Otsuka, M. Honma, T. Mizusaki, \emph{et~al.}, Monte Carlo shell model for atomic nuclei, Progress in Particle and Nuclear Physics, 47, 319-400, (2001).

\bibitem{Popov2000}
S. G. Popov J. J. Carbajo V. K. lvanov G. L. Yoder, 2000. Thermophysical Properties of MOX and UO{$_2$} Fuels Including the Effects ofIrradiation. ORNL/TM-2000/351

\bibitem{Broeders2000}
C.H.M Broeders, E Kiefhaber,and H.W Wiese.Burning transuranium isotopes in thermal and fast reactors. Nuclear Engineering and Design, 202, Issues 2-3, 2000, 157-172

\bibitem{Yuan2016c}
Yuan Cenxi, Chen Shengli, Zou Yaoleu, Study of MA in MOX fuel. CORPHY-2016, Pekin, Aug. 23-26, 2016

\bibitem{Iwasaki2002}
Tomohiko Iwasaki, 2002. A Study Of Transmutation of Minor-Actinide in a Thermal Neutron Field of the Advanced Neutron Source, Progress in Nuclear Energy, 40, 481-488

\bibitem{Yuan2016w}
Cenxi Yuan, Xuming Wang, Shengli Chen, 2016. A Simple Formula for Local Burnup and Isotope Distributions Based on Approximately Constant Relative Reaction Rate. Science and Technology of Nuclear Installation, 2016, 6980547

\bibitem{Yuan2016ch}
Cenxi Yuan, Shengli Chen, Xuming Wang, 2016. Monte Carlo Study on Radial Burnup and Isotope Distribution. PBNC 2016 Beijing, China, April 5-9, 2016

\bibitem{Yuan2016u}
Cenxi Yuan, Uncertainty decomposition method and its application to the liquid drop model, Physical Review C, 93, 034310 (2016)


\end{thebibliography}

\end{document}